\documentclass[showpacs,floats,aps,twoside]{revtex4}
\usepackage{amsmath,amssymb}
\usepackage{fancyhdr,lastpage}
 \usepackage{graphicx}
\usepackage{epsfig}
\usepackage{setspace}
\usepackage{float}
\floatstyle{plaintop}
\restylefloat{table}
 
\begin{document}

\title{Insulator -- half metallic transition by the tetragonal distortion: \\
A first -- principles study of strain -- induced perovskite RbMnF$_3$}

\author{Namsrai Tsogbadrakh$^{1}$}
\email{tsogbadrakh@num.edu.mn}
\author{N. Tuvjargal$^{1}$}
\author{Chun Feng$^{1,2}$}
\author{J. Davaasambuu$^1$}
\author{O. Tegus$^{2}$}
\affiliation{$^1$Department of Physics, Natural Science Division, School of Arts and Sciences, National University of Mongolia,
Ulaanbaatar 14201, Mongolia
\\
$^2$Inner Mongolia Key Laboratory for Physics and Chemistry of Functional Materials, 
Inner Mongolia Normal University, Hohhot 010022, China
}

\begin{abstract}
From the spin polarized density functional total energy calculations, we shown that the ground state of cubic perovskite Rubidium Trifluoromangate (RbMnF$_3$) is an antiferromagnetic (AFM) insulator due to the super -–exchange mechanism, in agreement with the other theoretical and experimental results. As included tetragonal distortion along the c -– axis, keeping the predicted volume, our results indicated that strain -–induced magnetic phase transition from an AFM insulator to a half metallic ferromagnetic (HM–FM) state is available by the tetragonal distortion due to the Jahn -- Teller distortion. We have shown that the easy axis of cubic perovskite RbMnF$_3$ is changed from the [111] direction to the [001] direction, as created the strain -– induced perovskite RbMnF$_3$ without any external magnetic field. The predicted electronic and magnetic properties of strain -– induced RbMnF$_3$ show the HM -– FM nature, making strain -– induced RbMnF$_3$ suitable for spintronic application.          
\end{abstract}


\maketitle           
\thispagestyle{plain}
\setcounter{page}{1}
\goodbreak

\section{Introduction}

Although the most common perovskite compounds contain oxygen, there are few perovskite compounds that form without oxygen. The perovskite oxide (e.g., BiFeO$_3$, BaTiO$_3$, SiTiO$_3$ etc) are known to undergo ferro -- or antiferro –- electric phase transitions, which are accompanied by distortion of the lattice to a lower crystallographic symmetry \citep{rb1}. Many of the alkali transition metal fluorides (e.g., KMnF$_3$, RbFeF$_3$, KCoF$_3$ etc) undergo similar phase transitions, which are apparently not associated with ferroelectric ordering \citep{rb2, rb3, rb4}. After the discovery of antiferromagnetism for rubidium trifluoromangate (RbMnF$_3$) \citep{rb5}, the elastic and magnetoelastic properties, nuclear acoustic resonance, magnetostriction and magnetocrystalline anisotropy (MCA) of single crystal AFM RbMnF$_3$ have experimentally investigated \citep{rb6, rb7, rb8, rb9, rb99}, and the magnetostriction and magnetoelastic couplings were measured at the 4.2 K in the magnetic fields up to 137 kOe \cite{rb9}. The theoretical and experimental investigations of Mn K -- edge for RbMnF$_3$ have shown the behaviors of 3d -- 4p intra -- atomic interaction in the conduction bands by resonant X -- ray magnetic scattering (RXMS) \citep{rb10, rb11}.    

In order to deeply understand the behavior of fluoroperovskite, the spintronic character and specially the magnetic properties of the RbMnF$_3$ compounds were studied by the first -- principles calculations \citep{rb13}. The AFM materials have been considering renewed attention due to the emerging materials of AFM spintronics \citep{rs1, rs2, rs3, rs4, rs5, rs6, rs7}. Commonly employed to pin the magnetization of an adjacent FM layer in spin valve devices through the interfacial exchange bias \citep{rs8, rs9, rs10}, among other developments AFM materials have been recently been shown to be efficient spin current detectors by meaning of the spin Hall effect \citep{rs11, rs12}. Three AFM materials that attracted considerable attention in the past are the following fluoride insulators: FeF$_2$, MnF$_2$ and RbMnF$_3$. These compounds show simple three dimensional AFM ordering with two sublattices at the temperatures below the Neel temperature of 78, 67 and 83 K respectively \citep{rs122}. Therefore, these insulators are not directly used to the AFM spintronics due to the low Neel temperatures, and are shown to be the paramagnetic (PM) phase at the room temperature. 

The magnetic interactions of FeF$_2$, MnF$_2$ and RbMnF$_3$ insulators are dominated by nearest neighbor exchange having effective exchange fields on the same order of magnitude 540, 515 and 830 kOe respectively \citep{rs122}. However, their magnetic anisotropy fields are different by several orders of magnitude. In FeF$_2$ compound, the ground state configuration of its magnetic Fe$^{2+}$ ions creates the $^5$D$_4$ term, which has a finite orbital angular momentum and consequently a large effective anisotropy field of 190 kOe, arising from the spin -- orbit coupling of single ion \citep{rs13, rs14}. But in  MnF$_2$ and RbMnF$_3$ compounds, the ground state configuration of their magnetic Mn$^{2+}$ ions creates the $^6$S$_{5/2} $ term with no single ion angular momentum, so that their crystalline anisotropy is small. In MnF$_2$, the tetragonal arrangement of the magnetic ions results in a sizable anisotropy of 10 kOe due to the  dipolar interaction \citep{rs15, rs16}. However, RbMnF$_3$ has a cubic perovskite structure with no measurable distortion, so that dipolar anisotropy vanishes. As this result, cubic perovskite RbMnF$_3$ has the very small magnetic anisotropy of 4.5 Oe \citep{rb5}. Recently Lopez Ortiz  was shown to occur the AFM -- spin flop (SF) and SF -- FM transitions due to the very small magnetic anisotropy, and obtained the critical magnetic field and SF temperature for the transition from the AFM phase to the SF phase \citep{rb99}.                             

In this study, we consider to occur the insulator -- half metallic transition of strain -- induced perovskite RbMnF$_3$ driving by tetragonal distortion by the first –- principles calculations within the framework of spin polarized density functional theory (DFT). Our results have shown the influence of tetragonal distortion to the magnetic property of strain -- induced perovskite RbMnF$_3$.   

\section{Computational Method} 

The RbMnF$_3$ usually crystallizes in the cubic perovskite structure with the space group of Pm -- 3m ($\#$ 221). The unit cell of RbMnF$_3$ has the 5 atoms and the atomic positions in RbMnF$_3$ are sited as follows: Rb atom at the (0, 0, 0), Mn atom at the (1/2, 1/2, 1/2), and F atoms at the (0, 1/2, 1/2), (1/2, 0, 1/2), (1/2, 1/2, 0). In order to create the AFM state, we used the (1 x 1 x 2) supercell in all the calculations. 

Our calculations are based on the pseudopotential projector augmented wave (PAW) and plane wave (PW) self -- consistent field methods using the generalized gradient approximation (GGA) by Perdew, Burke and Ernzerhof (PBE) \citep{rb14} within the framework of DFT \citep{rb15, rb16}, as implemented in the QUANTUM ESPRESSO 6.3 package \citep{rb17, rb18}. The interactions between the ions and valence electrons are expressed as the non relativistic ultrasoft \cite{rb19} and PAW \citep{rb1999} pseudopotentials taken from the Pslibrary 1.0.0 utility generated by A. Dal Corso \cite{rb199, rb19999}. The following electronic states are treated as valence states: Rb(4s$^2$, 4p$^6$, 5s$^1$), Mn(3s$^2$, 3p$^6$, 3d$^5$, 4s$^2$) and F(2s$^2$, 2p$^5$). The wave functions are expressed as plane waves up to a kinetic energy cutoff of 40 Ry and the kinetic energy cutoff for charge density and potential is chosen by 320 Ry. Three - dimensional Fast Fourier Transform (FFT) meshes for charge density, SCF potential and wavefunction FFT and smooth part of charge density are chosen to be (60 x 60 x 120) grids.  There might be need to use finer k -- points meshes for a better evaluation of on - site occupations due the strong correlated system. The summation of charge densities is carried out using the special k -- points restricted by the (10 x 10 x 5) grids of Monkhorst -- Pack scheme due to the computer power ability \citep{rb20}. The linear tetrahedral method is used when the electronic densities of state (DOS) are evaluated \citep{rb21}. To obtain optimized atomic structures, ionic positions and lattice parameters are fully relaxed until the residual forces are less than 0.05 eV/$\AA$ for each atom. The occupation numbers of electrons are expressed Gaussian distribution function with an electronic temperature of kT = 0.02 Ry. The mixing mode of charge density is chosen to be local density dependent Tomas -- Fermi (TF) screening for highly inhomogeneous systems. Its mixing factor for self -- consistency is to be 0.2 and the number of iterations used in mixing scheme is 5. The generalized eigenvalue problem is solved by the iterative diagonalization using the conjugate gradient (CG) minimization technique, and the starting wave function is chosen from superposition of atomic orbitals plus a superimposed "randomization" of atomic orbitals in all our calculation \citep{rb17, rb22}. In order to express the strong correlated effect of electrons in the Mn(3d) state, we first checked the U parameter of Hubbard -- based Hamiltonian on -- site Coulomb interaction from 2 eV to 7 eV, and was chosen to be U = 5 eV the using the simplified rotational -- invariant formulation based on the linear -- response method \citep{rb23}. Atomic wavefunctions used for GGA + U projector are not orthogonalized. In order to perform the MCA calculations \citep{rb233, rb2333, rb23333}, we have done the spin polarized density functional total energy calculations of non collinear magnetism (GGA + SOC) including the spin -- orbit coupling, using the fully relativistic ultrasoft and PAW pseudopotentials taken from the Pslibrary 1.0.0 utility.      

\section{Results and discussion }

We have first done the full relaxed total energy calculations of nonmagnetic (NM), FM and AFM states using both the PW and PAW methods by GGA and GGA + U approaches. We presented the results of predicted lattice parameters, band gap, magnetic energy gain between the FM and AFM states ($\Delta E = E_{FM} - E_{AFM}$), magnetic moments per atom and total magnetization of magnetic ions of RbMnF$_3$ on Table I. Our results are shown that the ground state of cubic perovskite RbMnF$_3$ is antiferromagnetically stable. In the PW and PAW methods, the AFM state is found to be energetically more stable by 24.49 and 26.72 meV/cell, respectively, than the FM one due to the super –- exchange mechanism through Mn -- F -- Mn bonding by the GGA approach. In these cases, the lattice parameter is predicted to be 4.16 $\AA$, and these values agree with the experimental values of 4.24 $\AA$ \citep{rb24}. The magnetic moment of Mn ion is found to be 4.69 $\mu_B$/atom by both the methods. But the band gaps are found to be 1.28 and 1.14 eV by the PW and PAW methods respectively. These values are different from the experimental value of 2.50 eV for cubic perovskite RbMnF$_3$ \cite{rb11}. Therefore, we considered the strong correlated effect of magnetic Mn ion by the U parameter of Hubbard -- based Hamiltonian on -- site Coulomb interaction. 

\begin{figure}[h]
\centering{\includegraphics[scale=0.55]{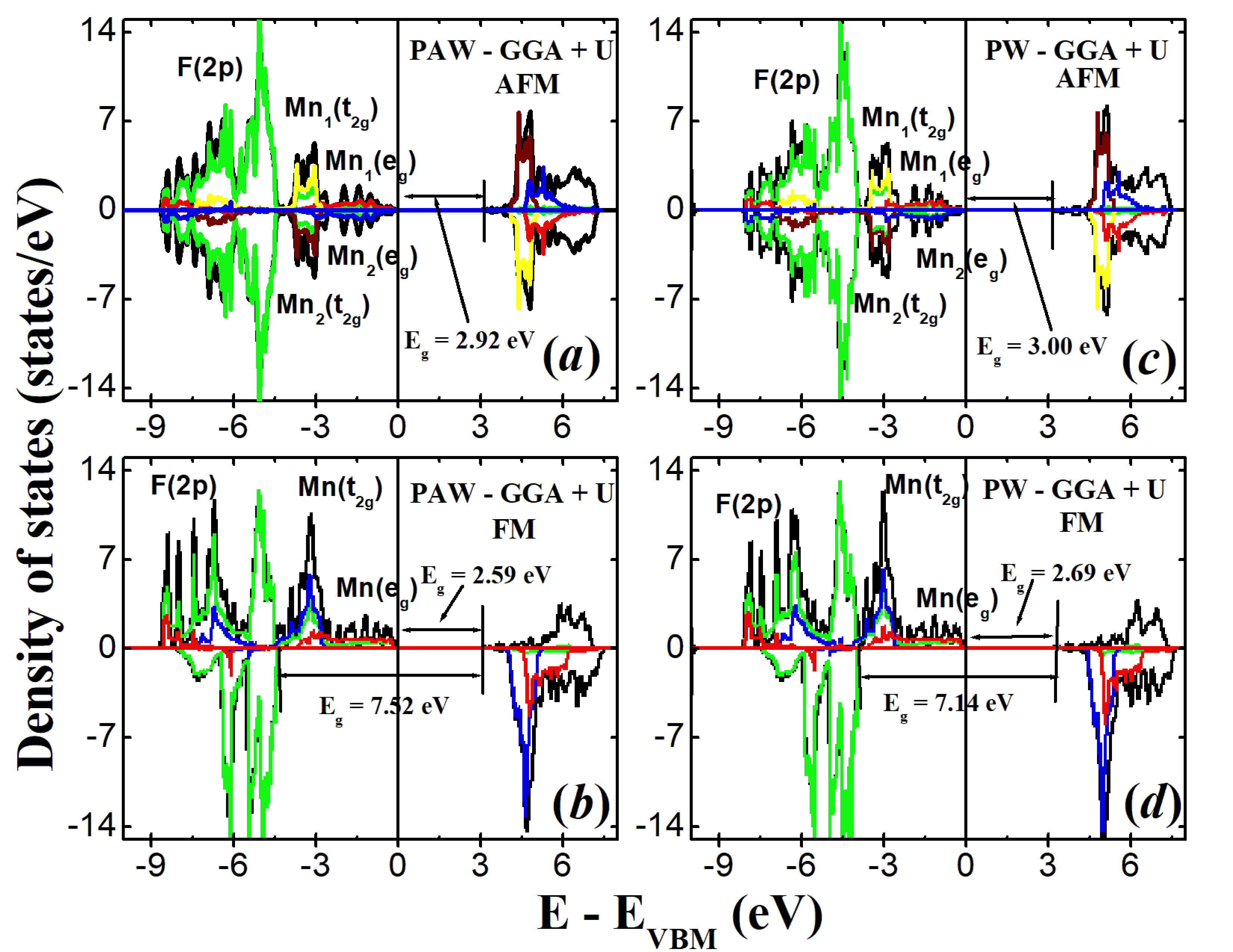}}
\caption{{ (Color online) The total and orbital projected electronic densities of states (TDOS and PDOS) of RbMnF$_3$ using the PW and PAW methods by the GGA + U approach. The valence band maximum (VBM) corresponds to the zero.
 }}\label{fig1}
 \end{figure}

For the GGA + U approach, the band gaps are found to be 3.00 and 2.92 eV by the PW and PAW methods respectively. It is shown an insulating behavior for both the majority and minority channels. We have shown the electronic total and orbital projected densities of state (TDOS and PDOS) of AFM and FM states for cubic perovskite RbMnF$_3$ using the PAW and PW methods by the GGA + U approach in the Figures (\ref{fig1}a, \ref{fig1}b) and (\ref{fig1}c, \ref{fig1}d) respectively. The electronic structure of GGA approach is similar to that of GGA + U approach. These values of band gap of AFM state for cubic perovskite RbMnF$_3$ agree with the other theoretical value of cubic perovskite RbMnF$_3$ \citep{rb11}. In this case, we note that the magnetic moments increase up to 4.79 and 4.77 $\mu_B$/atom by the PW and PAW methods respectively. The magnetic energy gains are found to be 6.04 and 11.16 meV/cell by the PW and PAW methods respectively. It is shown that the magnetic energy gain decreases, as included the strong correlated effect of magnetic Mn ion. These results affect to the lattice parameter and predicted lattice parameter decreases up to 4.10 and 4.04 $\AA$ by the PW and PAW methods respectively. 

For the AFM state, the Mn(3d) states are symmetrically and are splitting to the Mn(t$_{2g}$) and Mn(e$_g$) states by the octahedral crystal field of F ions. The F(2p) state is spreading from -8.6 eV to -2.4 eV and hybridized with the Mn(t$_{2g}$) state of Mn ion in both the PW and PAW methods. The main peaks of F(2p) states are sited at the positions of  -4.6 eV and -5.0 eV by the PW and PAW methods respectively. The Mn(e$_g$) state is located above the Mn(t$_{2g}$) state of Mn ion. For the unoccupied states, the separation of unoccupied Mn(t$_{2g}$) and Mn(e$_g$) states is to be smaller than that of occupied states. The intra -– atomic exchange splitting (Hund's coupling) is larger than the band gap of cubic perovskite RbMnF$_3$. For the FM state the majority and minority states are unbalancing and the band gap of minority state is increasing up to 7.14 and 7.52 eV by the PW and PAW methods respectively. These values are shown to be an insulating behavior for the minority channel. But the band gap of majority state is decreasing up to 2.69 and 2.59 eV by the PW and PAW methods respectively. It is shown a semiconducting behavior for the majority channel. These results agree with the theoretically results by Hashmi \citep{rb13}. 

For the PW and PAW methods by the GGA + SOC approach, the lattice parameter is increasing up to 4.30 $\AA$. The magnetic energy gains between the FM and AFM states are found to be 23.17 and 19.05 meV/cell, and the AFM state is favored to be the ground state of cubic perovskite RbMnF$_3$. The magnetic moments of magnetic ions is found to be 4.21 and 4.20 $\mu_B$/atom in the GGA + SOC approach by the PW and PAW methods respectively. These results are similar to the above results. While the GGA + SOC approach shows that the SOC is not small and it is affected to the lattice parameter of cubic perovskite RbMnF$_3$.          

The theoretical work is shown that the bulk modulus of cubic perovskite RbMnF$_3$ is smaller than that of other cubic perovskites RbXF$_3$ (X = V, Co and Fe) [13]. Therefore, we created the tetragonal distortion along the c -– axis to the cubic perovskite RbMnF$_3$ and the total energy calculations of NM, FM and AFM states have performed at the predicted volume. The magnetic energy gain and magnetic moments of Mn ion are shown in Figs. 2(a) and 2(b) by the GGA and GGA + U approaches, respectively. For the tetragonal distortion, when the ratio of c and a parameters becomes greater than 1.2 and 1.3 by the PW and PAW methods, respectively, the FM state is favored due to the insulator -– half metallic transition. We have shown the PDOS of Mn(3d) state in Figs. 2(c) and 2(d) using the PAW and PW methods by the GGA+U approach, respectively. The broadening of majority Mn (e$_g$) state for Mn ion is filling the majority band gap and crossing the Fermi level due to the Jahn-Teller distortion. The band gap of minority state is reduced up to 5.03 and 5.08 eV by the PAW and PW methods by the GGA+U approach, respectively. Therefore, the strain -– induced RbMnF$_3$ is shown a half metal behavior by tetragonal distortion due to the Jahn -- Teller distortion. The magnetic moment of Mn ion is decreasing up to 3.79 and 3.71 $\mu_B$/atom by the PW and PAW methods by the GGA + U approach, respectively. We have shown the TDOS of NM state for strain –- induced RbMnF$_3$ by the PAW and PW methods into the insets of Figs. 2(c) and 2(d), respectively. 

In our MCA calculations, we first estimated the MCE defined, as to be MCE=E$_{[100]/[010]/[001]}$ – E$_{[111]}$ in the cubic perovskite RbMnF$_3$. Our result is shown that the MCE is found to be 4.5 meV/cell. It is shown that the easy axis is located along the [111] direction and agrees with the experimental result of cubic perovskite RbMnF$_3$ [9]. For the HM -- FM strain -- induced perovskite RbMnF$_3$, we found to be the easy axis along the [001] direction, and the MCE’s of other spin orientations along the [100]/[010] and [111] directions are found to be higher than that along the [001] direction by 225 and 126 meV/cell, respectively. Therefore, we observed that the easy axis of cubic perovskite RbMnF$_3$ is changed from the [111] direction to the [001] direction without any magnetic field. 

In finally, we should note that all the results of the PAW methods by the GGA, GGA + U and GGA + SOC approaches are indicated to occur the AFM insulator -- half metallic ferromagnetic transition by the tetragonal distortion. It is shown that the HM -– FM state is favored by the Stoner mechanism of itinerant electrons. This behavior of strain –- induced RbMnF$_3$ show the HM -– FM nature, making strain -– induced RbMnF$_3$ suitable for spintronic application.

\section{ Conclusion}

In conclusion, we have predicted that the ground state of cubic perovskite RbMnF$_3$ is an AFM insulator due to the super -– exchange mechanism. As included tetragonal distortion along the c -– axis, keeping the predicted volume, our results indicated that the strain -– induced magnetic phase transition from an AFM insulator to a HM -– FM state occurs by the tetragonal distortion due to the Jahn -- Teller distortion. We observed that the easy axis of cubic perovskite RbMnF$_3$ is changed from the [111] direction to the [001] direction, as created the strain -– induced perovskite RbMnF$_3$ without any external magnetic field. The predicted electronic and magnetic properties of strain -– induced RbMnF$_3$ suitable for spintronic application. 

\begin{acknowledgments}
This work has supported by the research project of the Asia Research Center (Korean Foundation for Advanced Studies) titled “Study of rare -- earth magnetic materials” (code P2017 -- 1303) and Fundamental research project SSA$\_$014/2016 funded by the Mongolian Foundation for Science and Technology. We thanks for performing the calculations on the server computers at the School of Applied Science and Engineering and Nuclear Physics Research Center in the National University of Mongolia. 
\end{acknowledgments}

\newpage

\begin{table}
\begin{tabular}{cccccccc}
\hline \hline
& \multicolumn{3}{c}{PW} & &  \multicolumn{3}{c}{PAW}   \\ 
\cline{2-4} \cline{6-8}
 & \quad GGA\quad & \quad GGA + U\quad & \quad GGA + SOC \quad &  & \quad GGA \quad &\quad GGA + U \quad &\quad GGA + SOC  \\ 
\hline 
a($\AA$) & 4.16 & 4.10 & 4.30 & & 4.16 & 4.04 & 4.30  \\ 
$E_g$(eV) & 1.28 & 3.00 & &  & 1.14 & 2.92 &   \\ 
$\Delta E$(meV/cell) & 24.49 & 6.04 & 23.17 &  & 26.72 & 11.16 & 19.05      \\ 
M(Mn$_1$)($\mu_B$/atom) & 4,69 & 4.79 & 4.21 & & 4.49 & 4.77 & 4.20     \\ 
M(Mn$_2$)($\mu_B$/atom) & -4,69 & -4.79 & -4.21 & & -4.49 & -4.77 & -4.20   \\ 
M$_{tot}$($\mu_B$/cell) & 0.00 & 0.00 & 0.00 & & 0.00 & 0.00 & 0.00   \\ 
\hline\hline
\end{tabular} 
\caption{The predicted lattice parameters, band gap, magnetic energy gain between the FM and AFM states ($\Delta E = E_{FM} - E_{AFM}$), magnetic moments per atom and total magnetization of magnetic ions of RbMnF$_3$ using the PW and PAW methods by the GGA, GGA + U and GGA + SOC (U $= 5.0$ eV for Mn ion) approaches.}
\end{table}

\begin{figure}[h]
\centering{\includegraphics[scale=0.55]{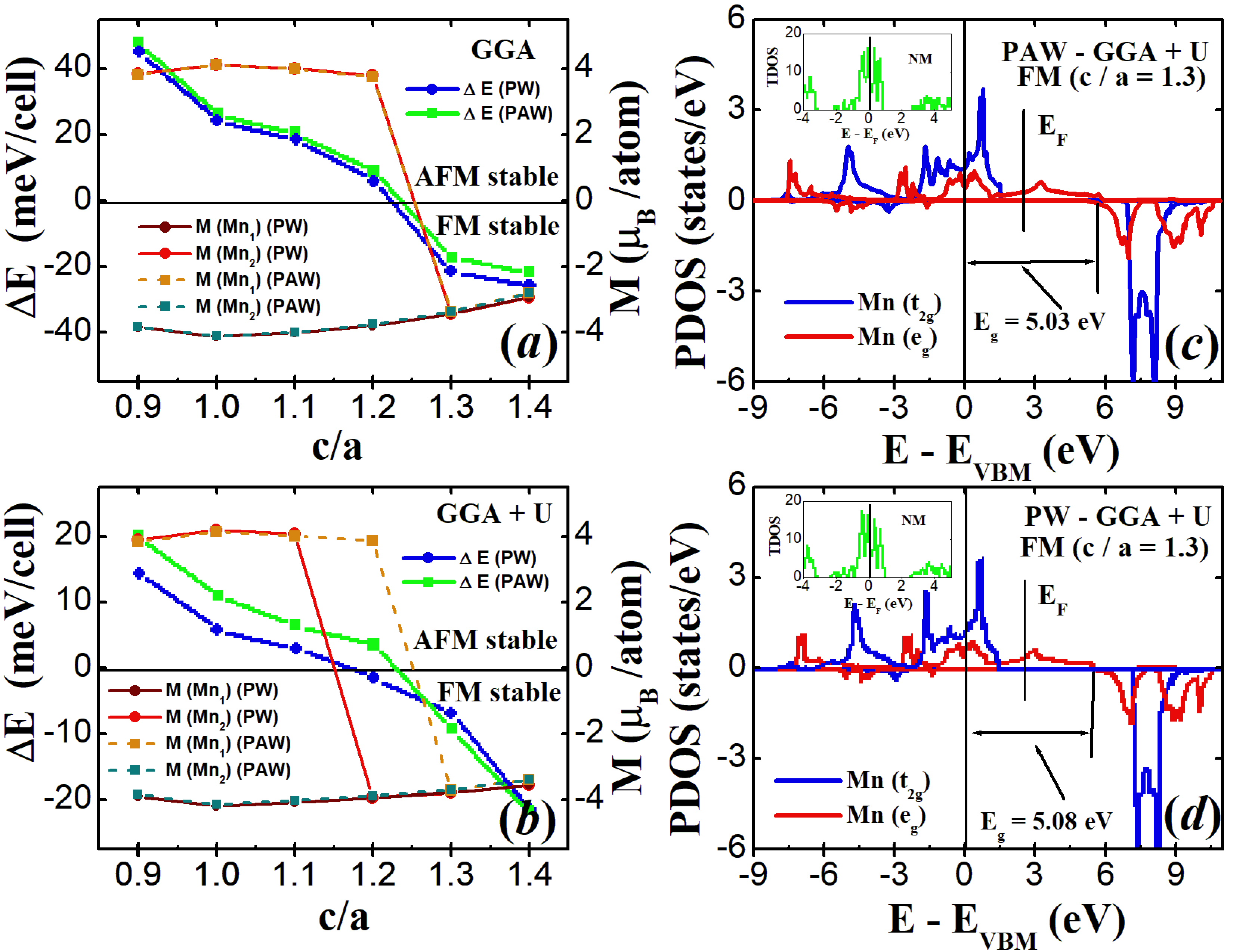}}
\caption{{ (Color online) The magnetic energy gains between the FM and AFM states ($\Delta$E = E$_{FM}$ - E$_{AFM}$) and magnetic moments of RbMnF$_3$ using the PW and PAW methods by the (a) GGA and (b) GGA + U approaches respectively. The orbital projected electronic densities of states (PDOS) of Mn (3d) state for strain -- induced perovskite RbMnF$_3$ (by the tetragonal distortion of c / a = 1.3) using the (c) PAW and (d) PW methods by the GGA + U approach respectively. The VBM of minority state corresponds to the zero. For the inserted figures, the total density of state (TDOS) of NM states for strain induced RbMnF$_3$ using the (c) PAW and (d) PW methods by the GGA + U approach respectively. The Fermi energy corresponds to the zero.  
 }}\label{fig2}
\end{figure}


\begin{thebibliography}{50}

\bibitem{rb1} ~	F. P. Jona and G. Shirane, 
\newblock Ferroelectric Crystals, Pergamon Press, New York, Chap. V (1962). 

\bibitem{rb2} ~	O. Beckman and K. Knox, 
\newblock Phys. Rev.  {\bf 121}, 376 (1961).

\bibitem{rb3} ~	L. R. Testardi, H. J. Levinstein and H. J. Gugenheim, 
\newblock Phys. Rev. Lett. {\bf 19}, 503 (1967), 

\bibitem{rb4} ~ J. D. Axe and G. D. Pettit,
\newblock Phys. Rev. {\bf 157}, 435 (1967). 

\bibitem{rb5} ~ D. T. Teaney, M. J. Freiser and R. W. H. Stevenson,
\newblock Phys. Rev. Lett. {\bf 9}, 212 (1962).

\bibitem{rb6} ~ R. L. Melcher and D. I. Bolef,
\newblock Phys. Rev. {\bf 178}, 178 (1969).

\bibitem{rb7} ~ R. L. Melcher and D. I. Bolef,
\newblock Phys. Rev. {\bf 186}, 491 (1969). 

\bibitem{rb8} ~ R. L. Melcher and D. I. Bolef,
\newblock Phys. Rev. {\bf 184}, 556 (1969).

\bibitem{rb9} ~	Y. Shapira and N. F. Oliveira Jr,
\newblock Phys. Rev. B {\bf 18}, 1425 (1978).

\bibitem{rb99} ~ J. C. Lopez Ortiz, G. A. Fonseca Guerra, F. L. A. Machado and S. M. Rezende, 
\newblock Phys. Rev. B {\bf 90}, 054402 (2014). 

\bibitem{rb10} ~ A. Stunault, F. de Bergevin, D. Wermeille, C. Vettier, Th. Bruckel, N. Bernhoeft, G. J. McIntyre and J. Y. Henry,
\newblock Phys. Rev. B {\bf 60}, 10170 (1999).

\bibitem{rb11} ~ M. Taguchi and M. Altarelli,
\newblock J. Electron Spectroscopy and Related Phenomena {\bf 136}, 205 (2004).

\bibitem{rb13} ~ M. R. Hashmi, M. Zafar, M. Shakil, A. Sattar, S. Ahmed and S. A. Ahmad,
\newblock Chin. Phys. B {\bf 25}, 117401 (2016).  

\bibitem{rs1} ~ A. S. Nunez, R. A. Duine, P. Haney and A. H. MacDonald,   
\newblock Phys. Rev. B {\bf 73}, 214426 (2006).

\bibitem{rs2} ~ A. B. Shick, S. Khmelevskyi, O. N. Miryasov, J. Wunderlich and T. Jungwirth, 
\newblock Phys. Rev. B {\bf 81}, 212409 (2010).

\bibitem{rs3} ~ A. H. MacDonald and M. Tsoi, 
\newblock Philos. Trans. R. Soc,, A {\bf 369}, 3098 (2011).

\bibitem{rs4} ~ V. M. T. S. Barthem, C. V. Colin, H. Mayaffre, M. -H. Julien and D. Givord,
\newblock Nat. Commun. {\bf 4}, 2892 (2013).

\bibitem{rs5} ~ P. Merodio, A. Ghost, C. Lemonias, E. Gautier, U. Ebels, M. Chshiev, H. Bea, V. Balz and W. E. Bailey
\newblock Appl. Phys. Lett. {\bf 104}, 032406 (2014).

\bibitem{rs6}  ~ C. Hahn, G. de Loubens, V. V. Naletov, J. B. Youssef, O. Klein and M. Viret,
\newblock arXiv: 1310.6000 (2013).

\bibitem{rs7} ~ E. V. Gomonnay and V. M. Loktev, 
\newblock Low Temp. Phys. {\bf 40}, 17 (2014).

\bibitem{rs8} ~ S. S. P. Parkin {\bf et. al.},
\newblock J. Appl. Phys. {\bf 85}, 5828 (1999).

\bibitem{rs9} ~ J. Nogues and I. K. Schuller, 
\newblock J. Magn. Magn. Mater. {\bf 192}, 203 (1999).

\bibitem{rs10} ~ J. R. Fermin, M. A. Lucena, A. Azevedo, F. M. de Aguiar and S. M. Rezende,  
\newblock J. Appl. Phys. {\bf 87}, 6421 (2000).

\bibitem{rs11} ~ H. Chen, Q. Niu and A. H. MacDonald,
\newblock Phys. Rev. Lett. {\bf 112}, 017205 (2014).

\bibitem{rs12} ~ J. B. S. Mendes, R. O. Cunha, O. Alves Santos, P. R. T. Ribeiro, F. L. A. Machado, R. L. Rodrigues -- Suarez, A. Azevedo and S. M. Rezende, 
\newblock Phys. Rev. B {\bf 89}, 140406(R) (2014).

\bibitem{rs122} L. J. de Jongh and A. R. Miedema,
\newblock Adv. Phys. {\bf 50}, 947 (2001).

\bibitem{rs13} ~ R. C. Ohlmann and M. Tinkham,
\newblock Phys. Rev. {\bf 123}, 425 (1961).
 
\bibitem{rs14} ~ M. T. Hutchings, B. D. Rainford and H. J. Guggenheim, 
\newblock J. Phys. C {\bf 3}, 307 (1970).

\bibitem{rs15} ~ F. Keffer,
\newblock Phys. Rev. {\bf 87}, 608 (1952).

\bibitem{rs16} ~ J. Barak, V. Jaccarino and S. M. Rezende, 
\newblock J. Magn. Magn. Mater. {\bf 9}, 323 (1978).

\bibitem{rb14} ~ J. P. Perdew, K. Burke and M. Ernzerhof,
\newblock Phys. Rev. Lett. {\bf 77}, 3865 (1996).

\bibitem{rb15} ~ P. Hohenberg and W. Kohn,
\newblock Phys. Rev. {\bf 136}, B864 (1964). 

\bibitem{rb16} ~ W. Kohn and L. J. Sham,
\newblock Phys. Rev. {\bf 140}, A1133 (1965).

\bibitem{rb17} ~ P. Gianmozzi, S. Baroni, N. Bonini, M. Calandra, R. Car, C. Cavazaaoni, D. Ceresoli, G. L. Chiarotti,  M. Cococcioni, I.Dabo, A. D. Corso, S. de Gironcoli, S. Fabris, G. Fratesi, R. Gebauer, U. Gerstmann, C. Gougoussis, A. Kokalj, M. Lazzeri, L. Martin-Samos, N. Marzari, F. Mauri, R. Mazzarello, S. Paolini, A. Pasquarello, L. Paulatto, C. Sbraccia, S. Scandolo, G. Sclauzero, A. P. Seitsonen, A. Smogunov, P. Umari and R. M. Wentzcovich,
\newblock J. Phys.: Condens. Matter {\bf 21}, 395502 (2009). 
   
\bibitem{rb18} ~ P. Giannozzi, O. Andreussi, T. Brumme, O. Bunau, M. Buongiorno Nardelli, M. Calandra, R. Car, C. Cavazzoni, D. Ceresoli, M. Cococcioni, N. Colonna, I. Carnimeo, A. Dal Corso, S. de Gironcoli, P. Delugas, R. A. DiStasio Jr, A. Ferretti, A. Floris, G. Fratesi, G. Fugallo, R. Gebauer, U. Gerstmann, F. Giustino, T. Gorni, J Jia, M. Kawamura, H.-Y. Ko, A. Kokalj, E. K¨u¸c¨ukbenli, M .Lazzeri, M. Marsili, N. Marzari, F. Mauri, N. L. Nguyen, H.-V. Nguyen, A. Otero -- de-la -- Roza, L. Paulatto, S. Ponc´e, D. Rocca, R. Sabatini, B. Santra, M. Schlipf, A. P. Seitsonen, A. Smogunov, I. Timrov, T. Thonhauser, P. Umari, N. Vast, X. Wu, S. Baroni,
\newblock J.Phys.: Condens. Matter {\bf 29}, 465901 (2017).

\bibitem{rb19} ~ D. Vanderbilt,
\newblock Phys. Rev. B {\bf 41}, R7892 (1990). 

\bibitem{rb1999} ~ P. Blochl,
\newblock Phys. Rev. B {\bf 50}, 17953 (1994). 

\bibitem{rb199} ~ Andrea Dal Corso,
\newblock Computational Materials Science {\bf 95}, 337 (2014). 

\bibitem{rb19999} ~ E. Kucukbenli, M. Monni, B. I. Adetunji, X. Ge,
G. A. Adebayo, N. Marzari, S. de Gironcoli, and A. Dal Corso,
\newblock arXiv:1404.3015v1 (2014). 

\bibitem{rb20} ~ H. J. Monkhorst and J. D. Pack,
\newblock  Phys. Rev. B {\bf 13}, 5188 (1976).

\bibitem{rb21} ~ P. E. Blochl, O. Jepsen and O. K. Andersen,
\newblock Phys. Rev. B {\bf 49}, 16223 (1994).

\bibitem{rb22} ~ M. C. Payne, M. P. Teter, D. C. Allan, T. A. Arias and J. D. Joannopoulos,
\newblock Rev. Mod. Phys. {\bf 64}, 1045 (1992).

\bibitem{rb23} ~ M. Cococcioni and S. de Gironcoli,
\newblock Phys. Rev. B {\bf 71}, 035105 (2005).

\bibitem{rb233} ~ G. Autes, C. Barreteau, D. Spanjaard and M. Desjonqueres, 
\newblock J. Phys.: Condens. Matter {\bf 18}, 6785 (2006).

\bibitem{rb2333} ~ D. Li, A. Smogunov, C. Barreteau, F. Ducastelle, and D. Spanjaard, 
\newblock Phys. Rev. B {\bf 88}, 214413 (2013).

\bibitem{rb23333} ~ X. B. Liu, Z. Altounian and D. H. Ryan,
\newblock J. Alloys and Compounds {\bf 688}, 188 (2016).

\bibitem{rb24} ~ R. L. Moreira and A. Dias,
\newblock J. Phys. Chem. Solids {\bf 68}, 1617 (2007).

\end{thebibliography}
\end{document}